# Origin of Electric Field Induced Magnetization in Multiferroic $HoMnO_3$

B. G. Ueland[1,*], J. W. Lynn[1], M. Laver[1,2], Y. J. Choi[3], S.-W. Cheong[3]

[1]*NIST Center for Neutron Research, National Institute of Standards and Technology, Gaithersburg, MD 20899, USA*

[2]*Department of Materials Science and Engineering, University of Maryland, College Park, MD 20742, USA*

[3]*Rutgers Center for Emergent Materials and Department of Physics & Astronomy, Rutgers University, Piscataway, NJ 08854, USA*

## Abstract

We have performed polarized and unpolarized small angle neutron scattering experiments on single crystals of $HoMnO_3$ and have found that an increase in magnetic scattering at low momentum transfers begins upon cooling through temperatures close to the spin reorientation transition at $T_{SR} \approx 40$ K. We attribute the increase to an uncompensated magnetization arising within antiferromagnetic domain walls. Polarized neutron scattering experiments performed while applying an electric field show that the field suppresses magnetic scattering below $T \approx 50$ K, indicating that the electric field affects the magnetization via the antiferromagnetic domain walls rather than through a change to the bulk magnetic order.

[*]*bgueland@nist.gov*

PACS numbers: 75.47.Lx, 75.80.+q, 75.60.Ch, 75.50.Ee

Multiferroics exhibit concurrent ferroelectric (FE) and magnetic order, and ones that show coupling between these two types of usually mutually exclusive order are of great scientific and technological interest [1]. Many members of the family of rare earth manganese oxides *Ln*$MnO_3$, where *Ln* represents a trivalent lanthanide cation, have shown multiferroic properties, and among them hexagonal $HoMnO_3$ is an intriguing multiferroic [2,3,4,5,6,7,8,9,10,11,12,13,14,15]. Specifically, anomalies in the dielectric susceptibility of $HoMnO_3$ occur at its magnetic transition temperatures indicating that coupling exists between the FE and magnetic order [5,9,11,16,17]. Perhaps most striking is that optical second harmonic generation (SHG) and Faraday rotation experiments have claimed that bulk ferromagnetic (FM) order of the Ho can be induced and reversibly altered through application of an electric field [4]. However, recent neutron and x-ray scattering experiments have found no evidence for any change in the bulk magnetic order while applying an electric field [18,19], warranting further study of this material. Here, we present results from small angle neutron scattering (SANS) experiments designed to explore the possibility that a net magnetization within antiferromagnetic (AFM) domain walls [20,21] maybe the origin of the previously observed electric field induced magnetization. We find that we can associate the SANS data with the magnetization arising within AFM domain walls and that a strong enough electric field affects this magnetic scattering.

$HoMnO_3$ has $P6_3cm$ symmetry in which $S = 2$ $Mn^{3+}$ form planes of side sharing triangles perpendicular to the c-axis. Looking down the c-axis axis, the Mn sublattices are spaced c/2 apart and rotated 60° with respect to one another, with $J = 8$ $Ho^{3+}$ ions also arranged in planes of side sharing triangles located between the Mn planes [22]. The effective interaction between Mn spins is AFM [2,9], and anisotropy keeps the spins constrained within the basal plane [2,8]. The magnetoelectric phase diagram shows that FE order, due to the displacement of Ho along the c-axis, occurs upon cooling through a temperature of $T_C = 875$ K, and AFM order occurs in the Mn sublattice at $T_N = 72$ K, with nearest neighbor spins pointing at 120° to one another [5,8,23,24]. Upon cooling through $T_N$, the overall symmetry changes to $P6'_3c'm$, followed by a subsequent change to $P6'_3cm'$ due to a magnetic field dependent spin reorientation transition occurring at $T_{SR} = 40$ K in zero magnetic field. [2,5,8,23,25,26]. At this transition, the Mn spins rotate 90° in the a-b plane but maintain a 120° relative orientation to their nearest neighbors. Below $T \approx 5$ K, another Mn spin reorientation transition occurs, which restores the $P6_3cm$ symmetry and is believed to be related to the magnetic order of the Ho sublattice [5,9,22,26]. For the Ho ions various descriptions of magnetic ordering of the Ho spins have been given [4,5,6,22,26]. Some reports claim that Ho spins on the 4b site develop AFM order below $T_{SR}$ while Ho spins on the 2a site remain paramagnetic [4,22], and other reports claim that all of the Ho spins develop AFM order below $T_{SR}$ [26]. However, there is general agreement that the Ho sublattice exhibits long range AFM order below $T \approx 5$ K [2,4,5,8,22,25,26,27].

Though there is not yet a consensus on the overall magnetic order of the Ho sublattice, previous results indicate that for $T < T_N$ the Ho sublattice can be driven FM through application of an electric field [4]. However, subsequent studies using magnetic x-ray and neutron scattering techniques have found no indication of field induced bulk FM order for fields up to $E = 300$ kV/cm [18,19], which greatly exceeds those used in the optical experiments ($E = 10^5$ V/cm) [4]. On the other hand, studies have pointed out the importance of coupling between AFM and FE domain walls in this and similar materials [5,27,28], and it is possible for a net magnetization to arise from uncompensated spins within AFM domain walls [21]. To shed light on this problem, we have performed SANS experiments while applying magnetic and electric fields. In the absence of ferromagnetic Bragg peaks, which would indicate bulk long range FM order, SANS allows one to examine any net magnetization occurring at small length scales, such as within AFM domain walls.

SANS experiments were performed on two different single crystal samples of $HoMnO_3$ grown by the floating zone method. One sample is cylindrical, approximately 5 mm in diameter and 15 mm long, and the second sample is thin and flat with dimensions 4 mm × 4 mm × 0.5 mm. These crystals are the same samples used in previous neutron experiments [2,8]. Data were taken on the NG3 and NG7 SANS spectrometers at the NIST Center for Neutron Research [29,30], using cold neutron beams with wavelengths of either $\lambda = 6$ Å or 8.4 Å and wavelength spreads of $\delta\lambda / \lambda = 0.34$ and 0.15, respectively. Measurements on the cylindrical sample were made in zero field and under magnetic fields up to $\mu_0 H = 8$ T applied perpendicular to the neutron beam. Measurements on the thin sample were made in zero field and while applying an electric field of $E = 25$ kV/cm along the c-axis, which was parallel to the neutron beam. For magnetic field measurements, data were taken after zero-field cooling from $T > 100$ K to 4 K in a superconducting magnet, while electric field measurements were taken in a closed cycle refrigerator after cooling from 150 K either with or without a field. Polarized SANS experiments were performed using a supermirror to polarize the incident neutron beam and a polarized $^{3}$He cell to analyze the polarization of the scattered neutrons [31,32]. Data were put on an absolute basis by normalizing to the incident flux when appropriate.

Figure 1 shows zero-field data for the cylindrical sample taken at various temperatures, using $\lambda$ = 6 Å neutrons. The total counts on the 2-D detector, normalized to the incident flux, are shown in Fig. 1a. The total counts increase with decreasing temperature, level off between 40 K $\leq T \leq$ 150 K and then increase more quickly with further temperature decrease. Plots in Fig. 1b were made after subtracting $T$ = 150 K data, to eliminate contributions from high temperature structural scattering, and then averaging over the 2-D SANS data to obtain the average scattering cross section, $[d\Sigma/d\Omega](|\mathbf{Q}|)$. (Uncertainties are within the size of the symbols unless otherwise indicated and are statistical in origin, representing one standard deviation.) In Fig. 1b we see that, for a specific temperature, $[d\Sigma/d\Omega]$ decreases with increasing $|\mathbf{Q}|$. Also, $[d\Sigma/d\Omega](|\mathbf{Q}|)$ is larger at lower temperatures, for low $|\mathbf{Q}|$. To illustrate this point, we plot $[d\Sigma/d\Omega](|\mathbf{Q}| = 0.0064$ Å$^{-1}$) as a function of temperature in Fig. 1c and see that $[d\Sigma/d\Omega]$ increases with decreasing temperature upon cooling below $T \approx 40$ K.

Data in the inset to Fig. 1b show that $[d\Sigma/d\Omega](|\mathbf{Q}|)$ follows a power law for at least $|\mathbf{Q}| < 0.012$ Å$^{-1}$ and fits to the data for each temperature yield a slope of -4. Hence, we fit data for each temperature over 0.0064 Å$^{-1}$ < $|\mathbf{Q}|$ < 0.012 Å$^{-1}$ to the Porod form,

$$|Q|^4 \frac{d\Sigma}{d\Omega}(|Q|) = \frac{4\pi\Delta\rho_m^2 S_m}{V}, \qquad (1)$$

which describes scattering from structures much larger than the length scale being measured [33]. In our case, this scattering results from structures with lengths much greater than $d = 2\pi / |\mathbf{Q}| \approx 100$ Å. In Eq. 1, $\Delta\rho_m$ is the contrast between different structures, $S_m$ is the surface area of the interface between structures, and $V$ is the sample volume [33,34]. The right hand side of Eq. 1 is the Porod amplitude and fits to the data are illustrated by lines in Fig. 1b. (We note that data taken at an order of magnitude lower in $|\mathbf{Q}|$ using a different instrumental configuration also follow the Porod Law.) In Fig. 1d, we plot the fitted Porod amplitude versus temperature and see that it is relatively constant at high temperatures but abruptly increases with decreasing temperature below $T \approx 40$ K. To test if this increase in scattering results from magnetic scattering, we performed polarized SANS experiments using a $\lambda = 8.4$ Å neutrons, and a plot of the total scattering for 0.0037 Å$^{-1}$ < $|\mathbf{Q}|$ < 0.013 Å$^{-1}$ versus temperature for the spin flip channel, which is sensitive only to magnetic scattering, is shown in the inset to Fig. 1a. No high temperature background subtraction is needed for the polarized data, and data were corrected for the $^3$He cell transmission, supermirror polarization, and spin flipper efficiency, resulting in a magnetic scattering cross section denoted as $\Sigma^{+-}$ [31]. Data in the inset show that a similar rise to the one observed in the main panel of Fig. 1a, 1c, and 1d occurs in $\Sigma^{+-}$ with decreasing temperature. These polarized beam data demonstrate conclusively that the abrupt increase in Fig. 1c and 1d below $T \approx 40$ K has a magnetic component. Thus, using the fact that bulk FM order does not occur in zero field in $HoMnO_3$, and that our experiments probe structures with $d > 100$ Å, we associate the observed increase in Porod scattering with decreasing temperature in Fig. 1d with magnetic domain walls that develop a net magnetization [5,20,21].

In order to test if our SANS measurements are indeed sensitive to changes to and the formation of magnetic domains, we performed SANS experiments on the cylindrical sample while applying a magnetic field and again fit the data to the Porod form. In Fig.

2a, we plot the change in SANS data with increasing magnetic field, $I(\mu_0 H)$, at various temperatures. These data were taken using $\lambda$ = 8.4 Å incident neutrons and are summed over 0.0020 Å$^{-1}$ < |**Q**| < 0.0065 Å$^{-1}$. We see that the initial slope of $I(\mu_0 H)$ increases with decreasing temperature, as is typical for field-induced scattering in magnetic systems. Furthermore, for $T$ = 4 K, $I(\mu_0 H)$ saturates above $\mu_0 H \approx 4$ T, similar to a canonical magnetization versus field curve. We also note that our data resemble previously published $M(\mu_0 H)$ data for $HoMnO_3$, which show magnetic saturation above $\mu_0 H \approx 4$ T at $T$ = 2 K [35]. In Fig. 2b we plot the Porod amplitude as a function of $\mu_0 H$ after performing fits over 0.0035 Å$^{-1}$ < |**Q**| < 0.022 Å$^{-1}$ to radially averaged $T$ = 4 K data taken at different magnetic fields. While the negligible change at low fields is likely due to a metamagnetic transition [2,5,8,23], the Porod amplitude increases with increasing field for $\mu_0 H > 1$ T until $\mu_0 H \approx 4$ T, where it saturates, similar to Fig. 2a.

Figure 3 shows unpolarized and polarized data for the thin sample taken while applying an electric field. In Fig. 3a, we show the temperature dependence of the total normalized 2-D detector counts, while applying $E$ = 25 kV/cm, after either field cooling or zero field cooling the sample. In the field cooled data, the counts increase with increasing temperature until $T \sim T_{SR}$, while for the zero field cooled data they decrease. Above $T_{SR}$ subsequent changes in the total counts with increasing temperature are qualitatively similar to $E$ = 0 data. The difference between the field cooled and zero field cooled data for $T < T_{SR}$ indicates that the electric field induces a macroscopic ground state different than that seen in Fig 1 for $E$ = 0, and that the induced ground state depends on the cooling protocol. Field cooled polarized SANS data are shown in Fig 3b. Here, we plot the spin flip scattering $\Sigma^{+-}$ as a function of temperature, after summing data over 0.004 Å$^{-1}$ < |**Q**| < 0.03 Å$^{-1}$, which indicate that the electric field affects the magnetic scattering in the same temperature range as in Fig. 3a.

Since it is well known that in the absence of a magnetic field AFM order exists in the Mn sublattice, and, when present, that the magnetic order in the Ho sublattice is likely AFM, then ideally no net magnetization is present in either sublattice. However, we have shown that the increase in low **Q** scattering with decreasing temperature below $T = 40$ K in Fig. 1 comes from magnetic structures, and that this magnetic scattering is affected by an electric field. Furthermore, the AFM order of either sublattice leaves open the possibility that the observed scattering results from net magnetization arising from uncompensated moments within AFM domain walls [5,20]. Using the interpretation of our fits to the Porod form to describe the data in Fig. 2, we postulate that the increase in magnetic scattering at low temperature for $E = 0$ in Fig. 1 is due to the development of a net magnetization within AFM domain walls. Thus, the increase in the Porod amplitude with decreasing temperature in Fig. 1d is due to either an increase in the contrast or number of AFM domain walls.

While AFM domains generally are not energetically favorable, effects such as lattice strain, defects, grain boundaries, or FE domains can lower the free energy [28,36,37]. Due to the unique nature of AFM domain walls, they can possess a net uncompensated magnetization, because they have lower symmetry than the bulk material [5,21,37]. Indeed, AFM domain walls have been observed in multiferroics [5,27,38], and pinning between FE domains and AFM domains in the Mn sublattice in both $YMnO_3$ and $HoMnO_3$ have been seen in recent experiments [27,28]. Since applying a strong enough electric field should change the FE domains, we can explain the observed changes in the magnetic SANS intensity upon application of an electric field in terms of pinned FE and AFM domains. When an electric field is applied, the change to the FE domains adjusts the AFM domains pinned to them, causing a change in the magnetization stemming from the walls of the pinned AFM domains. Interestingly, the data in Fig. 3b show that the electric field affects the intensity of the magnetic scattering for $T < T_{\mathrm{SR}}$, and the data in Fig. 1b show that for $E = 0$ the Porod amplitude increases with decreasing temperature for $T < 40$ K. This indicates that the electric field has the greatest affect on the magnetic scattering when changes are occurring to the zero field magnetic domain structure. We note that since some works report that the Ho sublattice possesses AFM order below $T_{SR}$ [4,22,26], our SANS data could also be the result of pinning between FE domains and AFM domains in the Ho sublattice. However, regardless of in which sublattice the magnetization originates, our data indicate that the previously observed electric field induced magnetization may originate from uncompensated spins in AFM domain walls, rather than conventional long range magnetic order. Nevertheless, the coupling of the electric field to magnetic domain walls is just as interesting and could prove useful in device applications, if such coupling occurs in a material for an appropriate temperature range.

## ACKNOWLEDGEMENTS

We gratefully acknowledge helpful discussions with O. P. Vajk, P. Butler, P. M. Gehring, D. Phelan, Y. Chen, and W. Ratcliff-II. B.G.U. acknowledges support from the NRC/NIST Postdoctoral Associateship Program. Work at the NCNR is supported in part by the NSF under Agreement No. DMR-0454672, and work at Rutgers is supported by the NSF under Agreement No. DMR-0520471.

FIG 1. (a) Temperature dependence of the normalized total counts. The inset shows the spin flip scattering cross section determined from the sum of the radially averaged polarized SANS data, as described in the text. (b) Radially averaged SANS data at various temperatures, after subtracting $T$ = 150 K data. The lines are fits to the Porod form. The inset shows data and fits plotted on a log-log scale. (c) Temperature dependence of data in (b) for $Q$ = 0.0064 $Å^{-1}$. (d) The Porod amplitude versus temperature as determined from the fits shown in (b). Lines are guides to the eye.

FIG 2. (a) Change in SANS intensity with an applied magnetic field at various temperatures. Data are shown after subtracting the zero field data at each temperature. (b) The Porod amplitude versus magnetic field at $T$ = 4 K. Lines are guides to the eye.

FIG 3. (a) Normalized total SANS intensity for warming with E = 25 kV/cm, after either field cooling (squares) or zero-field cooling (circles). (b) Field cooled spin flip scattering cross section versus temperature upon warming for $E$ = 25 kV/cm, as described in the text. Lines are guides to the eye.

Ueland *et al*. FIG. 1

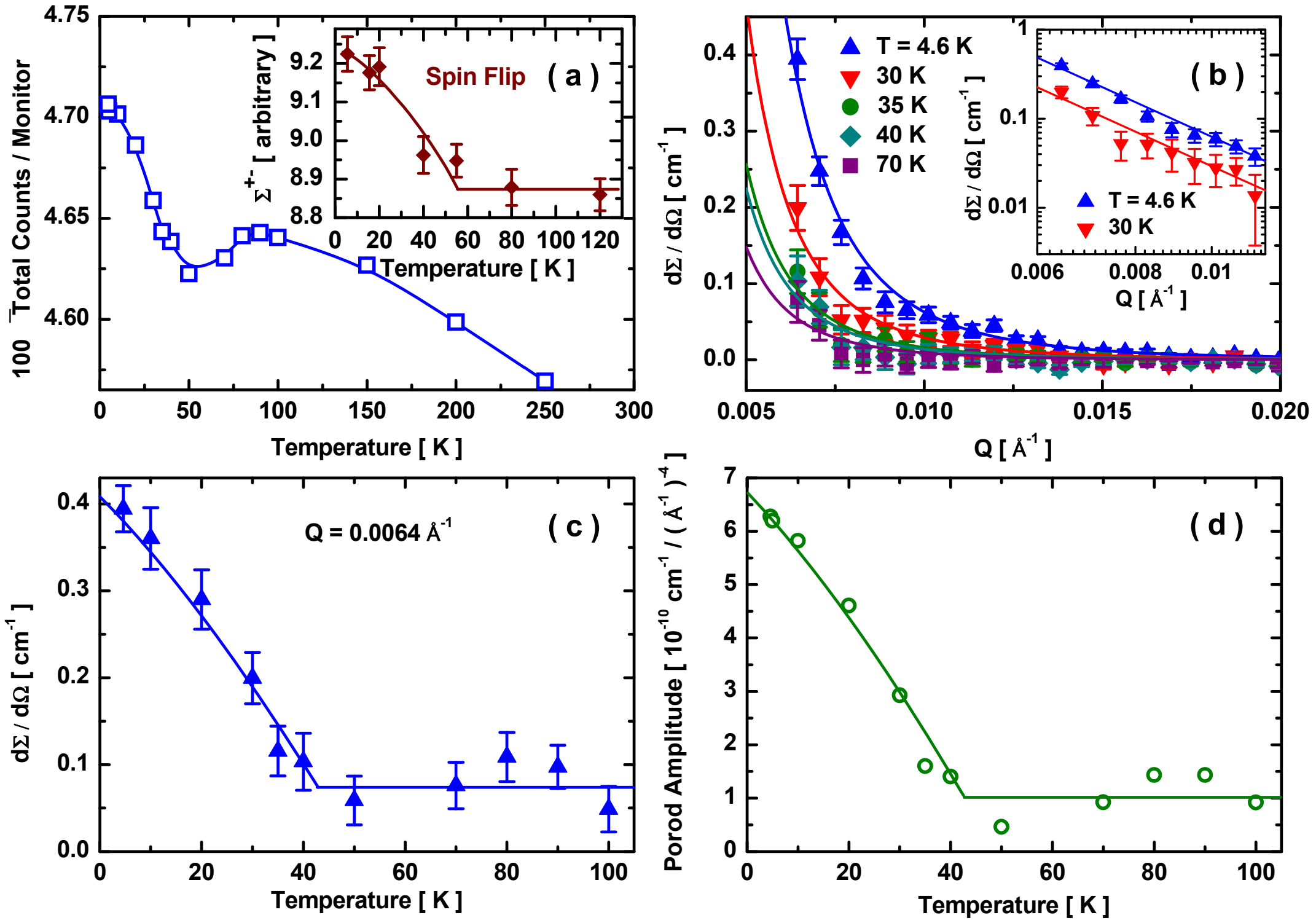

Ueland *et al*. FIG. 2

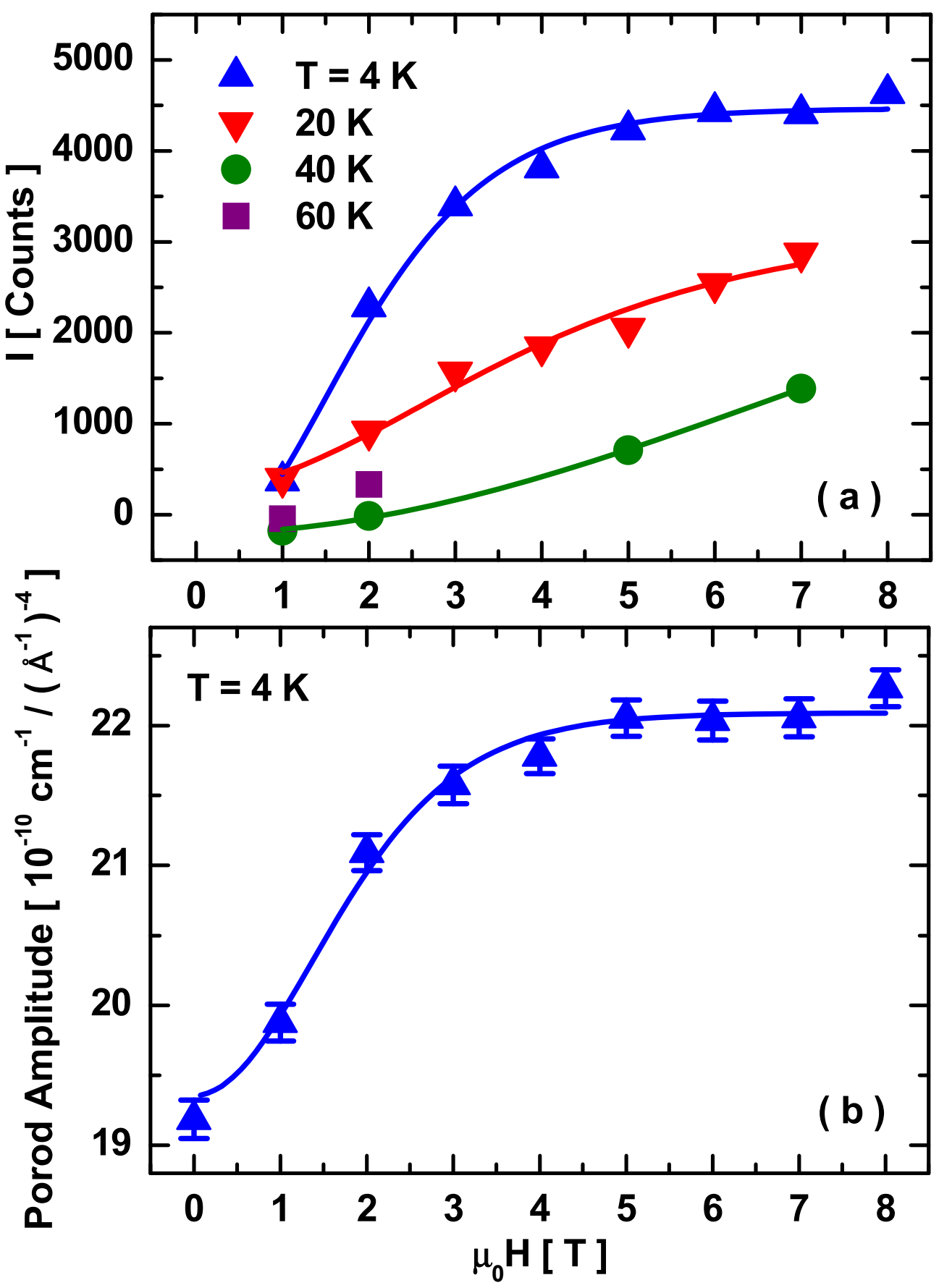

Ueland *et al*. FIG. 3

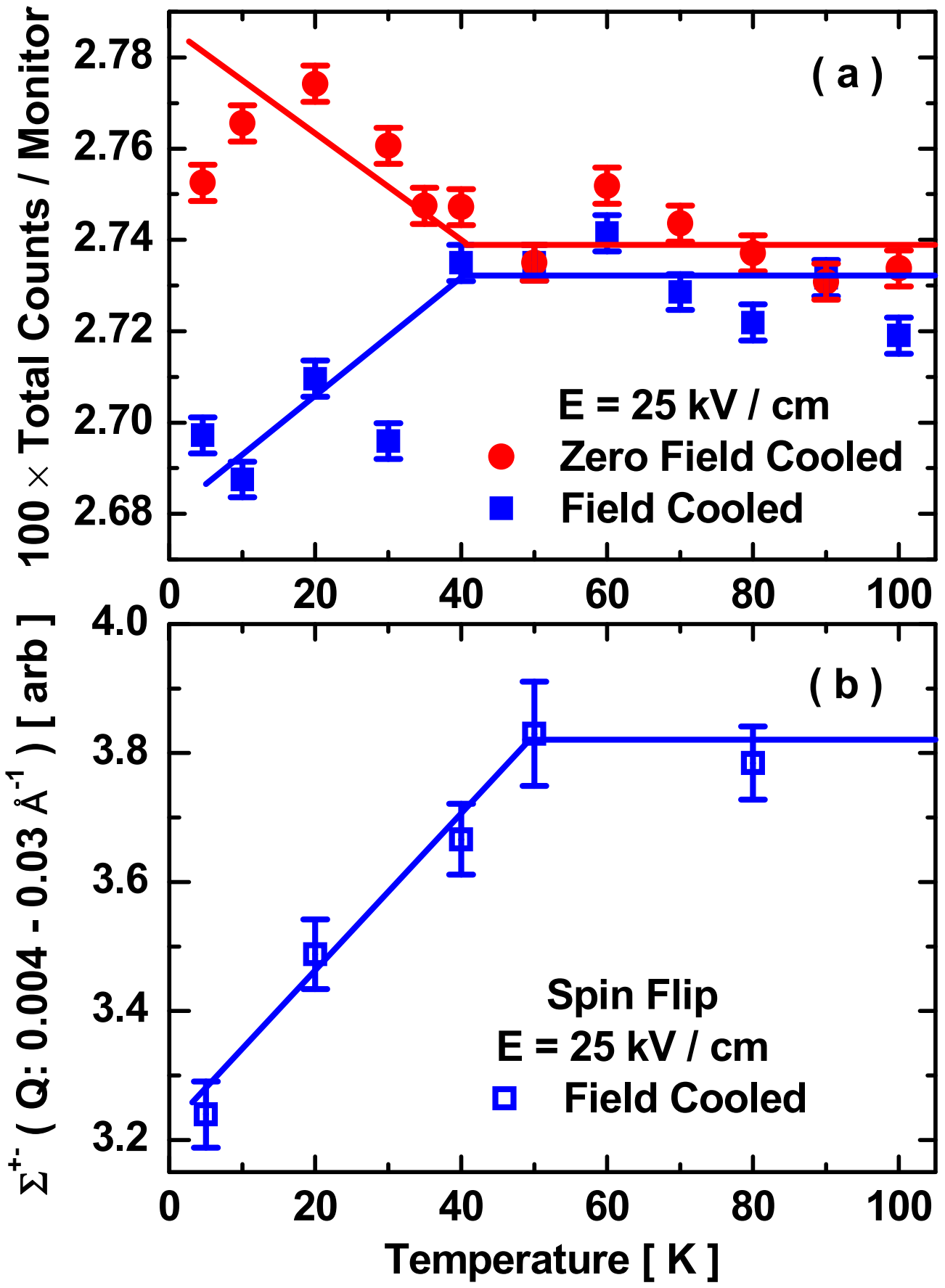